\documentclass[pre,twocolumn,showpacs]{revtex4}

\usepackage{graphicx}

\usepackage{amsmath,amssymb}
\usepackage{bm}
\usepackage{cancel}
\usepackage[normalem]{ulem}
\usepackage{hyperref}

\newcommand{\PR}{Phys. Rev. }
\newcommand{\PRL}{Phys. Rev. Lett. }
\newcommand{\JMP}{J. Math. Phys. }
\newcommand{\JPA}{J. Phys. A: Math. Gen. }
\newcommand{\EPL}{Europhys. Lett. }

\newcommand{\eq}{\text{eq}}

\begin{document}
\title{The Kovacs effect in the one-dimensional Ising model: a linear response analysis}
\author{M. Ruiz-Garc\'{\i}a$^1$ and A.\ Prados$^2$}
\affiliation {$^1$G. Mill\'an Institute, Fluid Dynamics, Nanoscience and Industrial
Mathematics, Universidad Carlos III de Madrid, 28911 Legan\'es, Spain}
\affiliation{$^2$  F\'{\i}sica Te\'{o}rica, Universidad de Sevilla,
Apartado de Correos 1065, E-41080 Sevilla, Spain, EU}

\date{\today}
\begin{abstract}
    We analyze the so-called Kovacs effect in the one-dimensional Ising model with Glauber dynamics. We consider small enough temperature jumps, for which a linear response theory has been recently derived. Within this theory, the Kovacs hump is directly related to the monotonic relaxation function of the energy. The analytical results are compared with extensive Monte Carlo simulations, and an excellent agreement is found. Remarkably, the position of the maximum in the Kovacs hump depends on the fact that the true asymptotic behavior of the relaxation function is different from the stretched exponential describing the relevant part of the relaxation at low temperatures.
\end{abstract}
\pacs{05.70.Ln,05.40.-a,81.05.Kf}
\maketitle

\section{Introduction}\label{intro}

Ever since the pioneering work of Kauzmann \cite{Ka48}, the interest in the investigation of the nature of the glassy state of supercooled liquids has steadily increased.  Real structural glasses have some characteristic behaviors, which are reproduced by many different models to a greater or lesser extent. A review thereof can be found in Refs.~\cite{Sch86,Sch90,ANMMyM00,ByB11}. Typically, the relaxation of the physical properties towards their equilibrium values at a given value of the temperature is non-exponential, being often well-fitted by the
stretched exponential or Kohlrausch-Williams-Watts (KWW) law \cite{Ko1854,WyW70}. When the supercooled liquid is cooled at a constant rate, there appears the phenomenon called the laboratory glass transition, in which the properties characterizing the macroscopic state of the system separate from their equilibrium values and eventually become frozen. This is due to the very fast increase of the typical relaxation times of the system
with decreasing temperature. Interestingly, when the system is reheated at the same rate, hysteresis effects are present: The system  overshoots the equilibrium curve and returns thereto only for higher temperatures. Therefore, the difference between the actual value of the macroscopic property of interest and its equilibrium value shows a non-monotonic behavior. In this way, there appears a memory effect in the system: its behavior depends on its whole thermal treatment, and not only on the instantaneous value of the property under consideration.

\begin{figure}
\begin{center}
\includegraphics[width=3.25in]{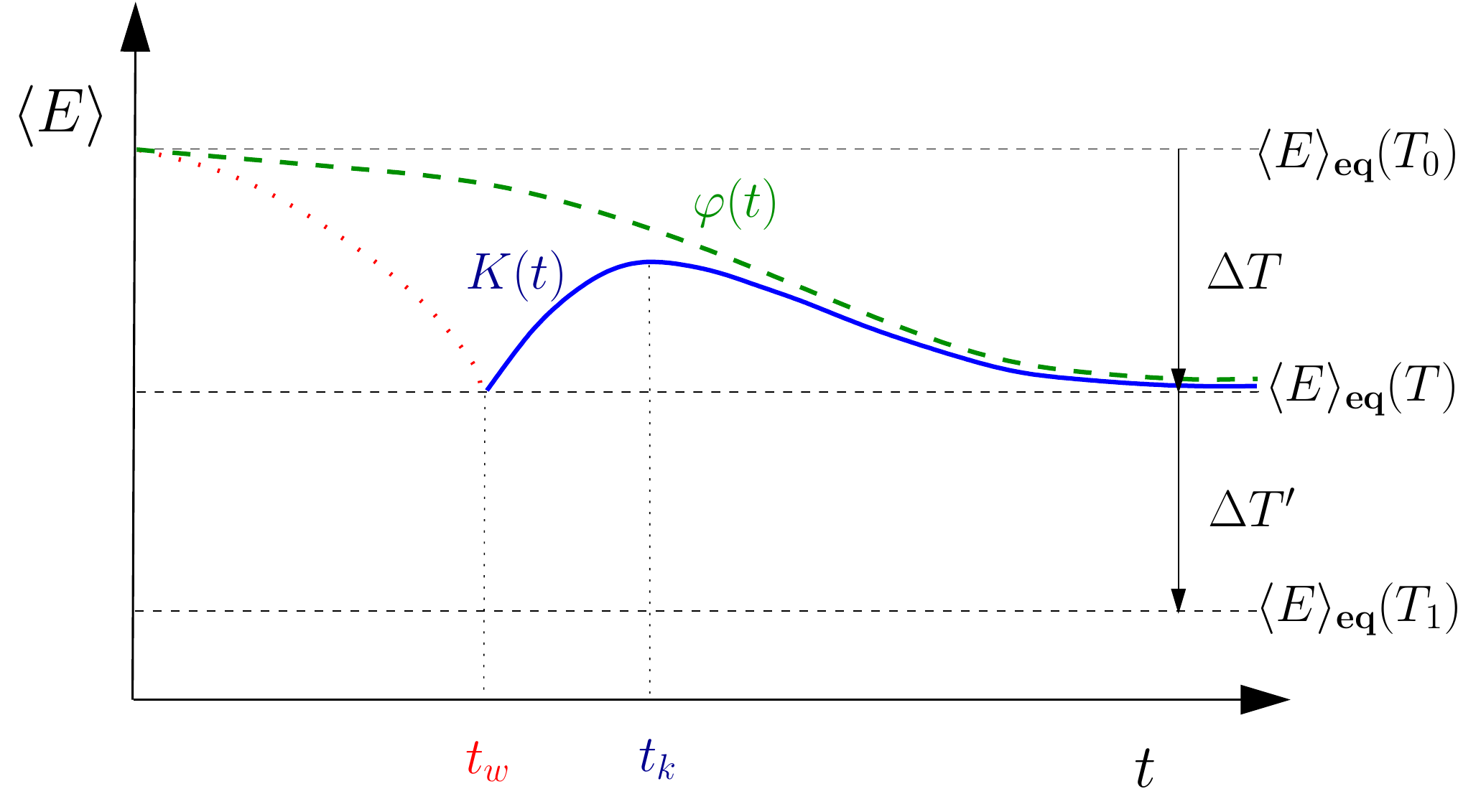}
\caption[]{Schematic representation of the Kovacs experiment described in the text. The dashed green curve $\varphi(t)$ represents the direct relaxation from $T_0$ to $T$. The dotted red curve stands for the part of the relaxation from $T_0$ to $T_1$, which is interrupted by the second temperature jump, changing abruptly the temperature from $T_1$ to $T$ at $t=t_w$. After this second jump, the system follows the solid blue curve $K(t)$, which reaches a maximum for $t=t_k$ and, afterwards, approaches $\varphi(t)$ for very long times.\label{fig1}}
\end{center}
\end{figure}

Here we focus on the memory effect that was first investigated by Kovacs \cite{Ko63,Ko79}, and thenceforth called the Kovacs effect. A sketch of the experimental procedure followed by Kovacs \cite{poly} is shown in Fig. \ref{fig1}, which starts from the equilibrium state corresponding to a high temperature $T_0$. First, an instantaneous quench to a lower temperature $T<T_0$ was done, and the direct relaxation of the energy to its equilibrium value $\langle E\rangle_{\eq}(T)$ was measured (curve $\varphi$). Secondly, a new program is started from equilibrium at $T_0$ but now the system is rapidly quenched to an even lower temperature $T_1<T<T_0$. The system then begins to relax to the equilibrium value of the energy at $T_1$, $\langle E\rangle_{\eq}(T_1)<\langle E\rangle_{\eq}(T)$. This relaxation is interrupted after a waiting time $t_w$  such that the instantaneous value of the energy $\langle E(t=t_w)\rangle$ equals $\langle E\rangle_{\eq}(T)$: At $t=t_w$, the temperature is suddenly increased to $T$. For $t>t_w$, the energy of the system does not remain flat, as one could naively expect. On the contrary, at first it increases, passes through a maximum at a certain time $t_k$, and finally returns to its equilibrium value. This simple experiment shows that, while the energy has its equilibrium value, the system is not actually in equilibrium at $t=t_w$. In fact, the subsequent evolution of the system depends on its previous thermal history. This statement is further supported by the behavior shown by the system when one fixes the temperatures $T_0$ and $T$, but the lowest temperature $T_1$ is changed. The maximum of the Kovacs hump function $K(t)$ increases as $T_1$ decreases or, equivalently, the temperature jump $T-T_1$ increases. Besides, the maximum moves to the left, in the sense that $s_k=t_k-t_w$ is a decreasing function of the jump $T-T_1$. Moreover, for very long times the Kovacs hump function $K(t)$ tends to approach the direct relaxation curve $\varphi(t)$.

A phenomenological description of this memory effect was given by Kovacs himself \cite{Ko79}. Also, the Kovacs effect has been extensively investigated, both analytically and numerically in several models \cite{Br78,MyS04,ByB02,CLyL04,Bu03,AyS04,ByH02,BBDyG03,TyV04,ALyN06,AAyN08,ByL10,DyH11,Be13}. However, it has not been until recently that a general theoretical expression of the Kovacs hump has been rigorously derived for systems whose mesoscopic dynamics is described by a master equation \cite{PyB10}. In this work, it has been shown that the Kovacs hump function $K(t)$, defined as
\begin{equation}\label{1bis}
  K(t)=\frac{\langle E(t)\rangle-\langle E\rangle_{\eq}(T)}{\langle E\rangle_{\eq}(T_0)-\langle E\rangle_{\eq}(T)},
\end{equation}
is given by
\begin{equation}\label{1} K(t)=\frac{\varphi(t)-\varphi(t_w)\varphi(t-t_w)}{1-\varphi(t_w)} \end{equation}
where the \textit{direct linear relaxation function} at temperature $T$ $\varphi(t)$ has been normalized, in the sense that $\varphi(t=0)=1$ and $\lim_{t\to\infty}\varphi(t)=0$ (see Sec. \ref{model}). It must be stressed that Eq. (\ref{1}) is a linear response theory result, nonlinear terms in the temperature jumps were neglected in its derivation. A great part of the typical behavior observed in the experiments is a direct consequence of the mathematical structure of (\ref{1}) and the non-exponential character of the direct relaxation. More concretely, it was analytically shown in Ref.~\cite{PyB10} that (i) $K(t)$ has only one maximum and that is always bounded by $\varphi(t)$, $0\leq K(t)\leq \varphi(t)$, (ii) the position of the maximum $s_k=t_k-t_w$ is a decreasing function of the second temperature jump $T-T_1$, and (iii) the Kovacs function tends to the direct relaxation function for very long times.

One of the simplest models that may be used to mimic complex systems is the one-dimensional Ising model with nearest neighbor interactions and Glauber dynamics \cite{Gl63}. Despite its simplicity, this system shows the main characteristic behaviors of structural glasses: the relaxation function of the energy is strongly non-exponential for low temperatures \cite{ByP93,ByP96}, the system displays a laboratory glass transition in which its energy becomes frozen when cooled down to low temperatures \cite{Sc88,ByP94}, and it exhibits a strong hysteretic behavior when reheated again to high temperatures, with a sharp peak of the apparent specific heat \cite{ByP94,BPyR94}. Moreover, aging is present for very low temperatures \cite{PByS97,GyL00}. The Kovacs effect in the Ising model was analyzed by Brawer long ago \cite{Br78}, but we would like to revisit it in light of the linear response results that we have already mentioned. We also investigate the behavior of the position and height of the maximum as a function of the temperature jump, relating them to the different stages of the direct relaxation for low temperatures.

The plan of the paper is as follows. We present the model in Sec.~\ref{model}, reviewing briefly its main linear response results. Section \ref{hump} is devoted to the analysis of the Kovacs hump. We compare simulation results to the analytical predictions. An excellent agreement between them is found, provided that the temperature jumps are small enough. We also explore non-linear effects, by considering larger temperature jumps.  Finally, the main conclusions of the paper are discussed in Sec.\ \ref{conc}.

\section{Model. Linear response results}\label{model}

We analyze the one-dimensional Ising model with Glauber dynamics \cite{Gl63}. We give only the main details that are
needed for understanding the work presented here. We have $N$ spins $\sigma_i=\pm1$, on a one-dimensional lattice of
$N$ sites labelled by $i=1,\ldots,N$.  The spins interact only with their nearest neighbors, with a ferromagnetic
coupling $J$. Thus, the energy of any configuration $\bm{\sigma}=\{\sigma_1,\ldots,\sigma_N\}$ is given by \begin{equation}\label{2} E(\bm{\sigma})=-J\sum_{i=1}^N \sigma_i \sigma_{i+1}. \end{equation}
We consider periodic boundary conditions throughout our work, so  we have that $\sigma_{N+1}=\sigma_1$ in the previous sum. The system is in contact with a heat bath at temperature $T$, and thus the dynamics of the model is stochastic and modelled in the following way: The probability distribution $p(\bm{\sigma},t)$ of finding the system in configuration $\bm{\sigma}$ at time $t$ obeys the master equation
\begin{equation}\label{3} \frac{d}{dt} p(\bm{\sigma},t)=\sum_{i=1}^N \left[ W_i(R_i\bm{\sigma}) p(R_i\bm{\sigma},t)-W_i(\bm{\sigma})p(\bm{\sigma},t)\right],
\end{equation}
where $R_i\bm{\sigma}$ is the configuration obtained from $\bm{\sigma}$ by rotating the $i$-th spin, $R_i\bm{\sigma}=\{\ldots,\sigma_{i-1},-\sigma_i,\sigma_{i+1},\ldots\}$, and $W_i(\bm{\sigma})$ are the transition rates for the Glauber single-spin-flip dynamics
\begin{equation}\label{4} W_i(\bm{\sigma})=\frac{\alpha}{2}\left[1-\frac{\gamma}{2}\sigma_i \left(\sigma_{i-1}+\sigma_{i+1}\right)\right].
\end{equation}
In the following, we set $\alpha=1$ without loss of generality, that is, we choose $\alpha^{-1}$ as the unit of time. On the other hand, $\gamma$ depends on both the coupling constant $J$ and the bath temperature $T$, \begin{equation}\label{5}
\gamma=\tanh \left(\frac{2J}{k_B T}\right).
\end{equation}
The dynamics above is ergodic, in the sense that any two configurations are connected through a chain of transitions with non-zero probability, and thus the system relaxes to thermal equilibrium for any initial condition. We restrict ourselves to homogeneous situations, in which the initial condition is translationally invariant; Glauber dynamics as defined by (\ref{3})-(\ref{5}) preserves this invariance for all times. In particular, the spin correlations
\begin{equation}\label{6} C_n \equiv \langle \sigma_i \sigma_{i+n}\rangle \to C_n^\eq =\xi^n, \qquad \xi=\tanh \left(\frac{J}{k_B T}\right), \end{equation}
in the long time limit.
\begin{figure}
\centering{\includegraphics[width=3in]{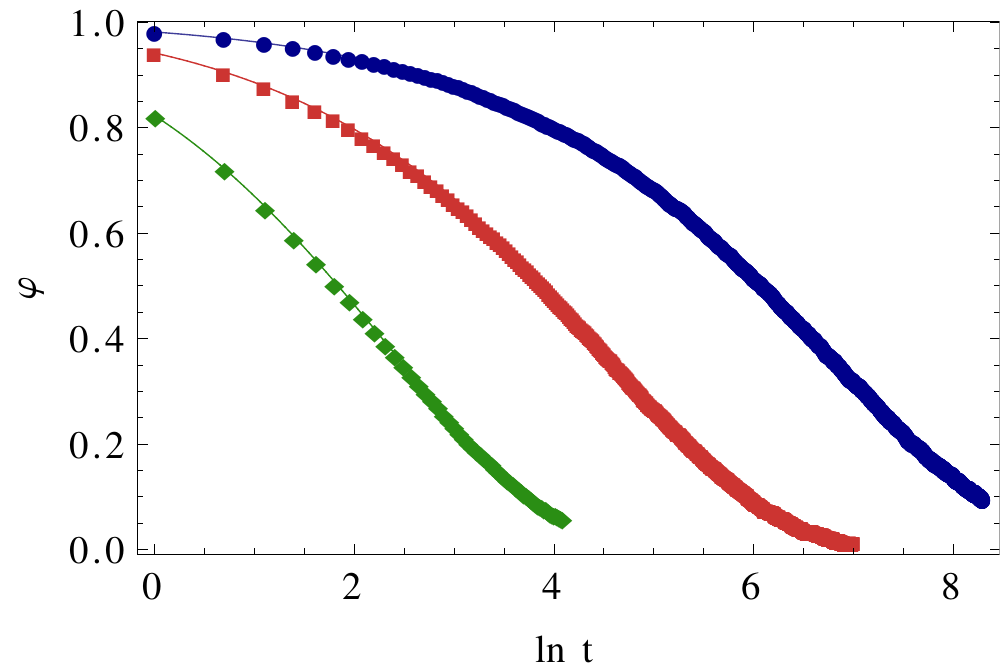}
            \vspace{0.5ex}
            \includegraphics[width=3in]{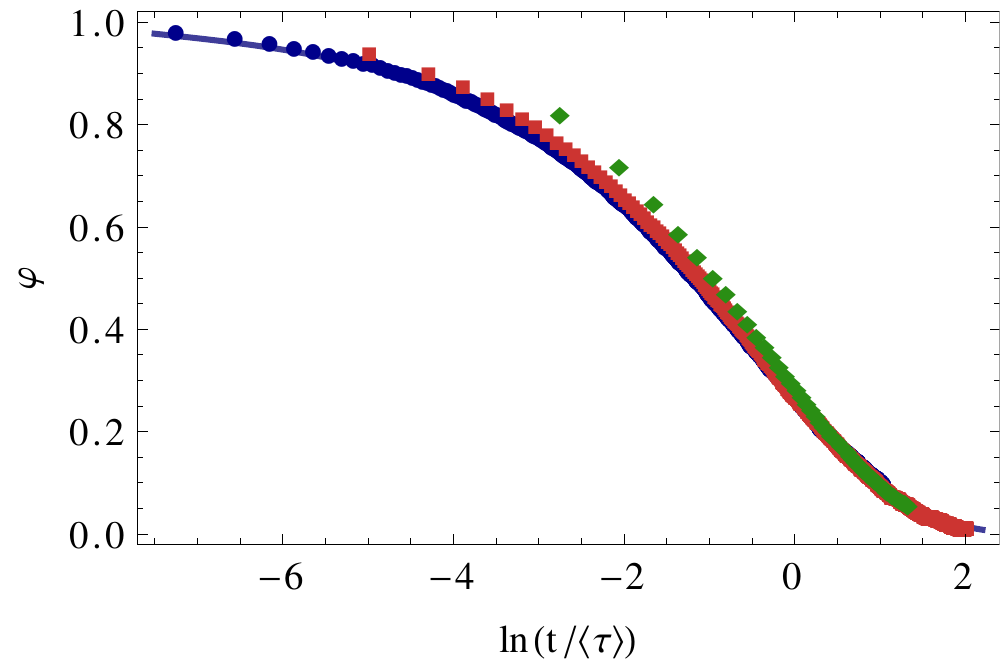}}
         \caption{\label{fig2} (Top panel) Comparison between the numerical evaluation (dots) and the analytical expression (solid line) of the direct relaxation function $\varphi(t)$ for the Ising model. The temperature jumps are: $\gamma=0.99 \to 0.991$ (green diamonds), $\gamma=0.999 \to 0.9991$ (red squares), and  $\gamma=0.9999 \to 0.99991$ (blue circles). Each numerical curve has been averaged over $10^6$ different realizations of the dynamics, using a system of $10^4$ spins. (Bottom  panel) Scaling property of the relaxation function. The relaxation function collapses when plotted versus the scaled time $t/\langle\tau\rangle$, for the same values of $\gamma$ as in the top panel figure. The thick solid line is the limit behavior for very low temperatures.}
\end{figure}

Let us consider the relaxation of the energy to its equilibrium value at a given temperature $T$, from  an initial state corresponding to equilibrium at temperature $T_0=T+\Delta T$, see Fig. \ref{fig1}. At time $t$, the system energy is denoted by
$\langle E(t)\rangle$, and the direct relaxation function characterizing this experiment is usually defined as
\begin{equation}\label{7}
  \varphi(t)=\frac{\langle E(t)\rangle-\langle E(\infty)\rangle}{\langle E(0)\rangle-\langle E(\infty)\rangle}=\frac{\langle E(t)\rangle-\langle E\rangle_\eq (T)}{\langle E\rangle_\eq(T_0)-\langle E\rangle_\eq(T)},
\end{equation}
so that $\varphi(0)=1$ and $\varphi(\infty)=0$. In the linear response regime, the relaxation function $\varphi(t)$
is independent of the temperature jump $\Delta T$, being a monotonically decreasing function of time. It reads \cite{ByP93}
\begin{equation}\label{8}
  \varphi(t)=\frac{\zeta(t)}{\zeta(0)}, \quad \zeta(t)=\int_0^\pi dq \, \frac{\sin^2 q}{(1-\gamma\cos q)^2} e^{-2 t (1-\gamma\cos q)},
\end{equation}
which is valid for all times $t$ and any value of the final temperature $T$. In the low temperature limit, $k_B T\ll J$, the relaxation becomes very slow. In fact, the average relaxation time $\langle\tau\rangle$, given by the area below $\varphi(t)$, diverges since
\begin{equation}\label{8bis}
  \langle\tau\rangle \sim \frac{1}{8\epsilon}, \quad \epsilon\equiv1-\gamma\sim 2 e^{-\frac{4J}{k_B T}}\to 0^+.
\end{equation}
This shows that the energy  relaxes over a very slow timescale $\epsilon t$. Moreover, the following three relevant regimes appear \cite{ByP93},
\begin{equation}\label{9}
  \varphi(t)\simeq \left\{ \begin{array}{ll}
  \exp[-2(2\epsilon)^{1/2}t], & 2 t \ll 1,  \\
  \exp\left[-(32\epsilon t/\pi)^{1/2}\right],   & 1\ll 2 t\ll \epsilon^{-1}, \\
  \pi^{-1/2} (2\epsilon t)^{-3/2} \exp (-2\epsilon t), & 2 t\gg \epsilon^{-1}.
  \end{array}
  \right.
\end{equation}
In the intermediate time regime $1\ll 2 t\ll \epsilon^{-1}$, the relaxation has the stretched exponential or KWW form,
\begin{equation}\label{10}
  \varphi(t)=\exp\left[ -(t/\tau)^\beta\right],
\end{equation}
with
\begin{equation}\label{11}
  \beta=\frac{1}{2}, \quad \tau=\frac{\pi}{32\epsilon}\sim\frac{\pi}{64}e^{\frac{4J}{k_B T}},
\end{equation}
where we have made use of (\ref{8bis}) for $\epsilon$.  In experiments, the situation is often similar to
that of \eqref{9}: the relaxation function is well fitted by a KWW law in the relevant intermediate time window. The actual long time behavior may differ from the KWW law, since the normalized relaxation function $\varphi$ is very small and it is difficult to measure \cite{Zw85}.

The strongly non-exponential relaxation shown by the Ising
model makes it adequate  to investigate, at least qualitatively, glassy-like behavior \cite{ByP93,Sc88,ByP94,BPyR94,ByP96,PByS97,GyL00}.
It shares characteristics of the fragile and strong liquids near the glass transition \cite{ANMMyM00}:  the value of $\beta$ is typical of fragile liquids, while the KWW relaxation  time $\tau$ follows an
Arrhenius-like behavior, as shown by strong liquids. Note that the KWW relaxation time $\tau$ is proportional to the average relaxation time $\langle\tau\rangle$, $\tau=\pi\langle \tau\rangle/4\simeq 0.79\langle\tau\rangle$.

Over the original time scale $t$, the relaxation function has a very small decrease, of the order of $\epsilon^{1/2}$, which can thus be neglected, but makes necessary to consider quite small values of $\epsilon$.  The KWW holds within an intermediate time window $t_i < t < t_f$, corresponding to a range of values $\varphi_{i} > \varphi(t) > \varphi_{f}$, where $t_i$ and $t_f$ may be estimated by calculating the intersection of the KWW function with the short and long time exponentials in Eq. (\ref{9}). Thus, $t_i=4/\pi$ and $2\epsilon t_f = 1.57$, which correspond to $\varphi_{i}=\exp[-8(2\epsilon)^{1/2}/\pi]$ and $\varphi_{f}=0.06$, respectively. For $\epsilon=10^{-2}$, it is $\varphi_{i}=0.70$, while for $\epsilon=10^{-4}$ it increases to $\varphi_{i}=0.96$. Therefore, most of the relevant relaxation of the energy can be accurately fitted by a KWW function at low enough temperatures, $\epsilon\lesssim 10^{-4}$. Moreover, the relaxation function has a universal form, since $\varphi_i \to 1^-$ for $\epsilon\to 0^+$: If we plot $\varphi$ as a function of $\epsilon t$ or, equivalently, $t/\langle\tau\rangle$, the curves corresponding to different temperatures collapse.  This means that the Ising model is thermorheologically simple in the very low temperatures regime, once the initial exponential regime in \eqref{9} becomes negligible. For moderately low temperatures, such that \eqref{9} gives a good description of the relaxation, but the three stages contribute thereto, there is a kind of weak thermorheological simplicity, in the sense that the direct relaxation curves collapse for long enough times, $t\gtrsim t_i$.

In Figure \ref{fig2} (top panel), we compare the numerical computation of the relaxation function $\varphi(t)$ to the analytical expression (\ref{8}) for three different values of $\epsilon=1-\gamma$, of the form $\epsilon=10^{-k}$, with $k=2,3,4$.  We have obtained the relaxation function by doing the actual experiment of equilibrating the system at
temperature $T+\Delta T$ and then measuring the evolution of the energy after a sudden temperature jump to $T$. The usual numerical procedure for calculating the linear relaxation function is to measure an equivalent time
correlation function at equilibrium, as given by the fluctuation-dissipation theorem \cite{vk92}. However, in the Kovacs experiment we must follow the actual procedure and thus we had to estimate the size of the temperature jumps needed for obtaining good averages. It is worth noting that the temperature jumps $\Delta T$ are not so small, since the average
relaxation time changes by ten percent between the initial and final temperatures. We also check the collapse of $\varphi$ when
plotted as a function of $t/\langle\tau\rangle$ in Fig.~\ref{fig2} (bottom panel). It is clearly observed that the
direct relaxation functions collapse onto a universal behavior for long enough times. Moreover, this time window extends
to all times in the limit as $T\to 0^+$ or $\epsilon\to 0^+$, in which the first stage of the relaxation in Eq. (\ref{9}) becomes
negligible, as discussed above. This is patent since, over the scale of the figure, the relaxation curve for $\epsilon=10^{-4}$ is indistinguishable from the limit behavior of Eq. \eqref{8} for $T\to 0^+$.

\section{Analysis of the Kovacs hump}\label{hump}

Now we investigate the Kovacs effect in the light of the recently derived results in the linear response regime \cite{PyB10}. For our present purposes, it is better to write the Kovacs hump as a function of the time after the second temperature jump $s$. Thus, we rewrite Eq.~(\ref{1}) in the form
\begin{equation}\label{kovacs}
  K(s)=\frac{\varphi(t_w+s)-\varphi(t_w)\varphi(s)}{1-\varphi(t_w)}, \qquad s=t-t_w.
\end{equation}
We restrict ourselves to the low temperature limit, in which the Ising model shows glassy-like behavior, as discussed throughout the previous Section. We do the Kovacs experiment in the following way (see Fig. \ref{fig1}): (i) we consider fixed values of $T_0$ and $T_1$, such that the direct relaxation of the energy from $T_0$ to $T_1$ is accurately given by the linear response function  (\ref{8}) (ii) we consider different values of $t_w$ along this direct relaxation, which lead to different values of the temperature $T$ at which the Kovacs hump is measured. It should be noted that this is slightly different from the usual procedure in the experiments, in  which $T_0$ and $T$ are kept constant and different values of $T_1$ are considered. The problem with this procedure is that the second temperature jump $T-T_1$ is not bounded and will eventually become very large, making linear response theory not applicable. This is the reason why we have chosen fixed values of $T_0$ and $T_1$, the variation of the final temperature $T$ poses no problem. As discussed in the previous section,  the direct relaxation curves corresponding to different values of the final temperature $T$ collapse onto a unique behavior when plotted versus the rescaled time $t/\langle\tau\rangle$. This allows us to compare in one plot the Kovacs humps corresponding to different values of $t_w$.
\begin{figure}
  \centering
    \includegraphics[width=3.25in]{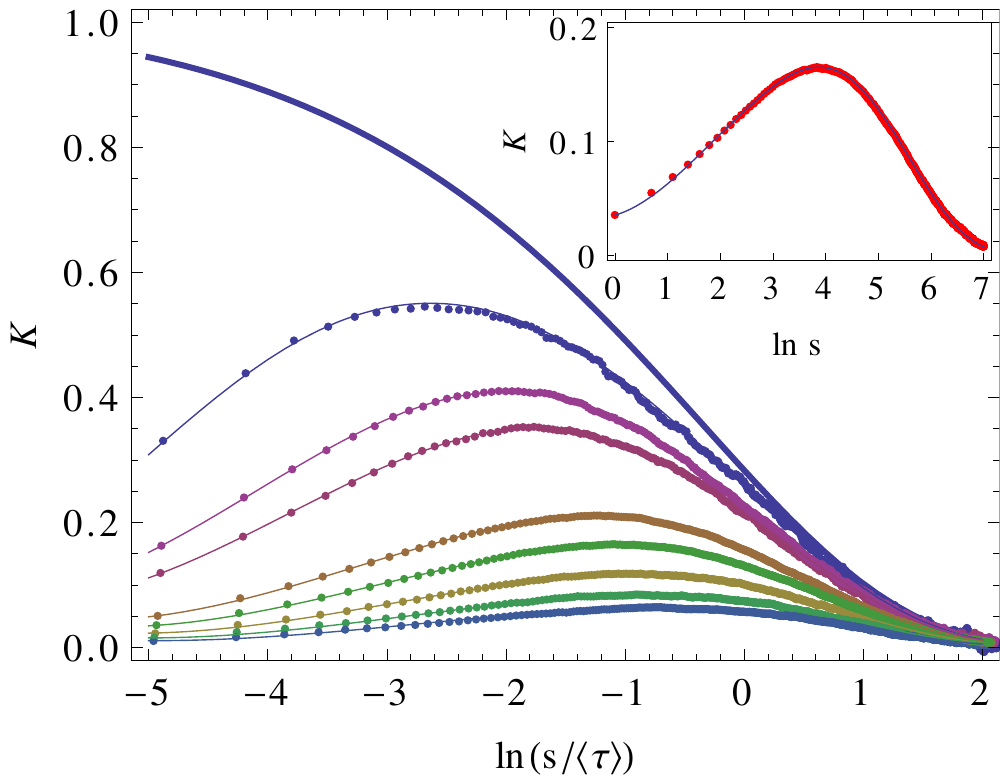}
           \caption{Comparison of the numerical evaluation and the theoretical prediction of the Kovacs hump. We show the numerical curves (points) and the theoretical predictions \eqref{kovacs} (line) as a function of the time from the second temperature jump $s=t-t_w$. In the main panel, we plot Kovacs hump for different values of the waiting time, namely $t_w=1$, $7$, $13$, $50$, $80$, $130$, $195$, and $260$ (from top to bottom). Note that the maximum increases and moves to the left as $t_w$ decreases and, at the same time, $K(s)$ tends to the direct relaxation
 function $\mathbf{\varphi}$ (upper thick solid line) for long enough times. A zoom of one of the curves, namely that of $t_w=80$, is plotted in the inset. It shows that the agreement between theory and simulation is really good, even if observed on a  much finer scale. In both graphs, $\gamma_0=0.999$  and $\gamma_1=0.9991$, which corresponds to a change of the relaxation time of around $10\%$.
}
  \label{fig3}
\end{figure}

In  Fig. \ref{fig3}, we plot the numerical evaluation of the Kovacs hump for a experiment in which $\gamma_0=0.999$ and $\gamma_1=0.9991$, together with the analytical prediction (\ref{kovacs}). In the inset, the waiting time is $t_w=80=0.6\langle\tau\rangle$,   where $\langle\tau\rangle$ is the average relaxation time for the temperature $T$ corresponding to this value of $t_w$  . The agreement between the numerical and the theoretical curve is excellent. In the main panel, several Kovacs humps for the same values of the extreme temperatures $\gamma_0$ and $\gamma_1$ but different values of the waiting time are plotted. As discussed above, they correspond to different temperatures $T$ and thus we plot them as a function of the rescaled time $s/\langle\tau\rangle$.  It is clearly seen that the behavior is completely similar to the experimentally observed one: the Kovacs hump $K(s)$ moves to the left and  its maximum increases as the waiting time decreases; moreover, $K(s)$ approaches the direct relaxation curve for long enough times. Again, the agreement between the simulation and the theoretical expression is excellent in all cases. The same is true for other values of the limiting temperatures $\gamma_0$ and $\gamma_1$, provided that the temperature jump is small enough to assure the validity of the linear response theory result \eqref{kovacs}. As a rule of thumb, temperature jumps corresponding to changes of the average relaxation time by ten per cent are still small enough.

\begin{figure}
  \centering
    \includegraphics[width=3.25in]{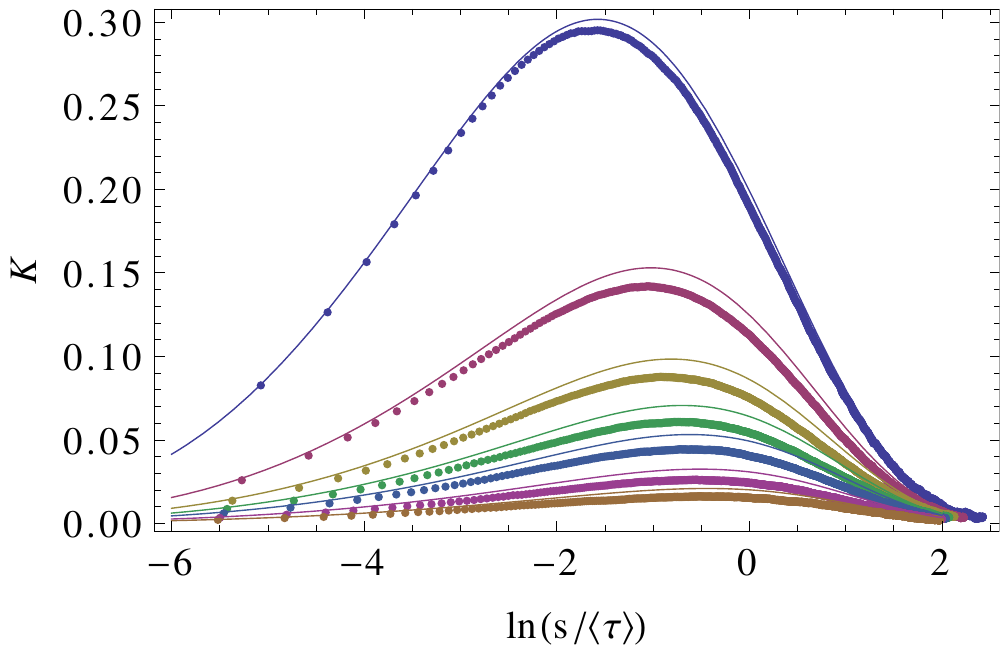}
\caption{\label{fig:non-linear}  Comparison of the numerical evaluation and the theoretical prediction of the Kovacs hump in the non-linear regime. The Kovacs hump function is plotted (points-simulation, line-theory) against the time from the second temperature jump $s$, scaled with the average relaxation time $\langle\tau\rangle$ at the final temperature $T$.  The temperature $T$ at which \eqref{kovacs} is evaluated is taken from the simulations. From top to bottom, the waiting times are $t_w=25, 128, 257, 386, 515, 773, 1031$.  Note that we use a quite large first temperature jump: $\gamma_0=0.999$ and $\gamma_1=0.9995$. 
}
\end{figure}

We explore the non-linear regime in Fig.~\ref{fig:non-linear}. The initial temperature is the same as in Fig.~\ref{fig3}, that is, $\gamma_0=0.999$, but a much lower limit temperature is set, namely $\gamma_1=0.9995$. Again, we consider different values of the waiting time $t_w$ leading to different intermediate temperatures $T$, and the measured
Kovacs hump is compared to the linear response expression. Of course, the quantitative agreement is not as good as in Fig.~\ref{fig3}, but the linear theory still gives a reasonable description of the observed hump. The estimate for the maximum  position remains very good,  but for the maximum height there appear some quantitative discrepancies: The relative error in its estimate, which increases with $t_w$ in the considered range, is always bounded by $25\%$. This is remarkable, since we are using a quite
large jump: the relaxation time at $\gamma_1$ roughly doubles that of $\gamma_0$. Recall that $\langle\tau\rangle\propto (1-\gamma)^{-1}$, as given by Eq.~\eqref{8bis}.

\subsection{Position and height of the maximum}

Let us analyze in more detail the behavior of the Kovacs hump. In particular, we will investigate the position and the height of the maximum as a function of the waiting time $t_w$. As most of the relaxation is
accurately given by the KWW function in (\ref{10}), we substitute it into Eq.~(\ref{kovacs}) and look for the value $s_k$ that makes $K'(s)$ vanish. Following Ref.~\cite{PyB10}, we introduce the definitions
\begin{equation}\label{11b}
  D_1(s)=-\frac{d}{ds}\ln\varphi(s),\quad D_2(s)=-\frac{d}{ds}\ln\varphi'(s).
\end{equation}
For the KWW function,
\begin{subequations}\label{12}
\begin{equation}\label{12a}
    D_1^{KWW}(s)=\frac{\beta}{\tau} \left(\frac{\tau}{s}\right)^{1-\beta},
\end{equation}
\begin{equation}\label{12b}
     D_2^{KWW}(s)=\frac{1-\beta}{s}+D_1^{KWW}(s).
\end{equation}
\end{subequations}

First we analyze the limit of small waiting times, $t_w\ll\langle\tau\rangle$, so that
\begin{equation}\label{13} \delta=1-\varphi(t_w)\ll 1.
\end{equation}
For $s\gg t_w$ (a time window that widens as $t_w$ goes to zero), the analytical expression \eqref{kovacs} is well approximated by \begin{equation}\label{14} K(s) \sim \varphi(s)+\frac{t_w}{\delta}\varphi'(s)=\varphi(s)\left[ 1-\frac{t_w}{\delta} D_1(s) \right].
\end{equation}
The position of the maximum is given by the solution of the equation $D_2(s_k)=\delta/t_w$ (equivalent to Eq.~(57) of Ref.~\cite{PyB10}). The first term on the rhs of $D_2^{KWW}$ is the dominant one ($s\gg t_w$), which makes it possible to give the estimate
\begin{equation}\label{15} \frac{s_k}{\tau} \sim (1-\beta) \left( \frac{t_w}{\tau} \right)^{1-\beta},
\end{equation}
consistently with the assumption $s\gg t_w$. The height of the maximum $K_{\max}$ is obtained by making use of (\ref{14}), \begin{equation}\label{16} \frac{K_{\max}}{\varphi(s_k)}=\frac{K(s_k)}{\varphi(s_k)}\sim 1-\frac{\beta}{1-\beta} \left( \frac{s_k}{\tau}\right)^\beta,
\end{equation}
where $s_k/\tau$ is given by Eq.~(\ref{15}). Moreover, Eq.~(\ref{14}) implies that
\begin{equation}\label{17} \lim_{s\to\infty} \frac{K(s)}{\varphi(s)}=1, \end{equation}
since $D_1^{KWW}(s)\ll \delta/t_w$ for long enough $s$. This agrees with the experimental observations and the  numerics of the model (see Fig.~\ref{fig3}): The Kovacs hump approaches the direct relaxation function for long enough times.
\begin{figure}
  \centering
\includegraphics[width=3.25in]{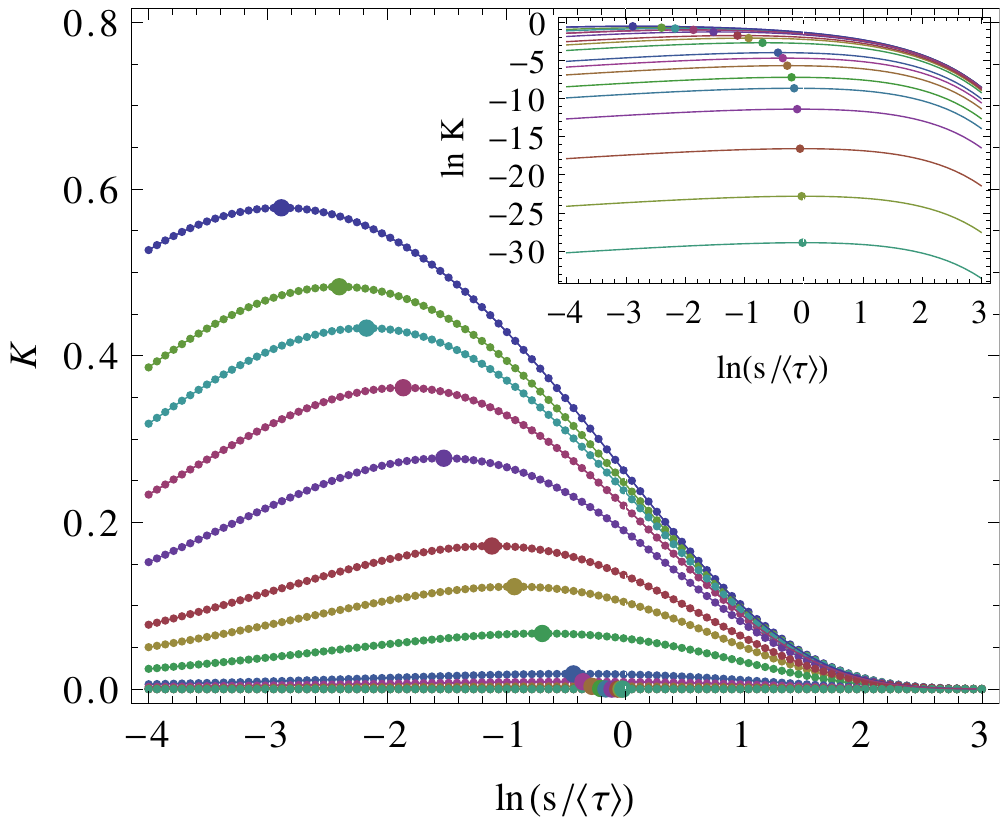}
\caption{\label{fig5} Theoretical Kovacs hump function (\ref{kovacs}) in the linear response approximation, for $\gamma=0.9999$,
and different values of the waiting time. The direct relaxation function has been evaluated by numerically integrating (\ref{8}),
and we have considered the waiting time values
$t_w= 13$, $40$, $65$, $130$, $280$, $760$, $1300$, $2500$, $6350$, $9000$, $13000$, $19000$, $26000$, $38000$, $64000$, $ 95000$,
$ 127000$ ($t_w/\langle\tau\rangle$ ranging from $0.01$ to $100$, from top to bottom). The position of the maximums are also
indicated therein (circles). Inset: The same plot but with  the vertical axis also in a logarithmic scale, in order to see
the trend for long waiting times more clearly.}
\label{max}
\end{figure}

Let us now investigate the opposite limit of long waiting times $t_w\gg\langle\tau\rangle$, such that $\varphi(t_w)\ll 1$.
 In Ref.~\cite{PyB10}, it was obtained that
\begin{equation}\label{18}
  K(s)\simeq \varphi(t_w) \left[ e^{-D_1(t_w)s}-\varphi(s) \right],
\end{equation}
provided that $D'_1(t_w)s^2\ll 1$ (which becomes always true for long enough times, because $D'_1$ tends to zero). If the KWW function were used to approximate both $\varphi$ and $D_1$, the position of the maximum would be given by
\begin{equation}\label{19}
  \frac{s_k}{\tau}\sim\left[ (1-\beta) \ln\left(\frac{t_w}{\tau}\right)\right]^{1/\beta}
\end{equation}
and $s_k$ would diverge logarithmically. As will be seen, this does not agree with the Monte Carlo simulations of the Ising model. This was to be expected since the KWW function is not valid for $t_w\gg\tau$: In this regime $\varphi$ is given approximately by the long time exponential (with
algebraic corrections) in Eq. (\ref{9}). Then, we look for a different dominant balance, by assuming that $s_k/\tau$ remains of
the order of unity for large waiting times. We thus use the long time exponential for $\varphi(t_w)$ and $D_1(t_w)$, but
the KWW function for $\varphi(s)$. In this way, the maximum of (\ref{18}) is given by the solution of the equation
\begin{equation}\label{20}
\beta {s_{k}^{*}}^{\beta-1} e^{ -{s_{k}^{*}} ^{\beta}+c s_{k}^{*}} \sim c, \quad c=\frac{\pi}{16}
\end{equation}
where $s_k^*=s_k/\tau$. Consistently with our assumptions, $s_k^*$ remains of the order of unity. The height of the Kovacs hump follows by substituting Eq.~(\ref{20}) into Eq.~(\ref{18}), with the result
\begin{equation}\label{21}
 K_{\max}=K(s_k)\sim \varphi _{E} (t_{w}) \left[e^{-2\epsilon s_k} - \varphi _{E} (s_{k})\right],
\end{equation}
where we have taken into account that $D_1(t_w) \to 2\epsilon$ for $t_w \gg \langle \tau \rangle$.

\begin{figure}
  \centering
\includegraphics[width=3in]{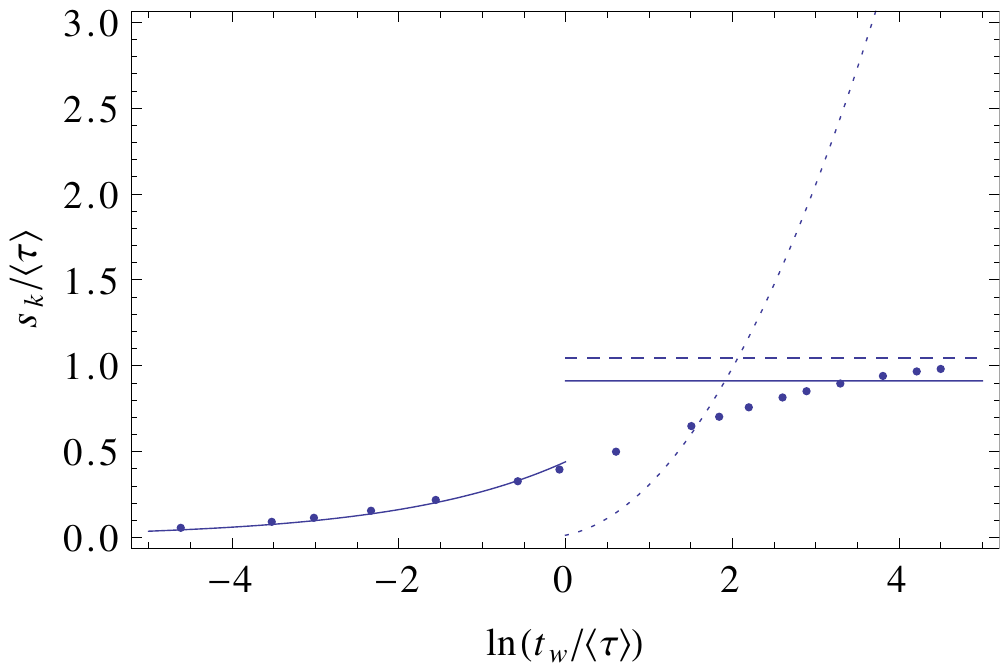}
\vspace{0.5ex}
\includegraphics[width=3in]{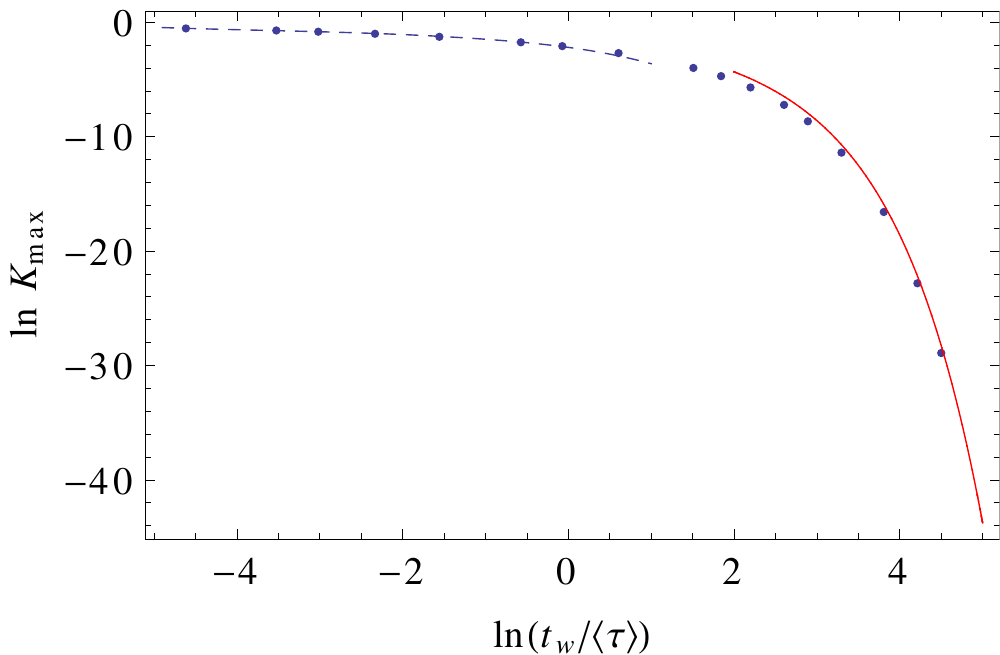}
    \caption{(Top panel) Position of the maximum of the Kovacs hump $s_k$ as a function of the scaled waiting time
$t_w/\langle\tau\rangle$. The points are the maximum positions in Figure \ref{max}, while the solid lines are the
theoretical predictions (\ref{15}) and (\ref{20}). The dashed line corresponds to the value of $s_k$ obtained by
substituting the values of $\beta$ and $\tau$ of the best KWW fit into (\ref{8}) instead of the theoretical values
(\ref{11}). The dotted line corresponds to the incorrect, logarithmically diverging, KWW prediction (\ref{19}) for
long waiting times. (Bottom panel) Height of the maximum of the Kovacs hump $K_{\max}$ as a function of the scaled
waiting time $t_w/\langle\tau\rangle$. The points are the heights of the maximums in Figure \ref{max}, while the
solid lines are the asymptotic expressions for short times (\ref{16}) (dashed blue) and for long times (\ref{21})
(solid red). In both graphs, $\gamma_0=0.9999$ and $\gamma_1=0.99991$. }\label{fig6}
\end{figure}

We have chosen $\gamma_0=0.9999$ and $\gamma_1=0.99991$ to  numerically study the behavior of the maximums of the Kovacs hump. We have already presented the direct relaxation between them in Fig.~\ref{fig2}. As discussed in Sec. \ref{model},  the initial exponential stage in \eqref{9} scarcely contributes to the relaxation ($\epsilon_0=1-\gamma_0=10^{-4}$), and the KWW extends to almost all the relevant part thereof. Figure \ref{fig5} shows the theoretical prediction for the Kovacs hump for different values of the waiting time, with the maximums indicated therein. Again, it is clearly observed that the maximum moves to the left and increases in height as the waiting time decreases (or, equivalently, the second temperature jump increases as compared to the direct relaxation one). Figure \ref{fig6} compares the maximum positions in Fig.~\ref{fig5} to our asymptotic estimates. A remarkably very good agreement is found, both for the position and height of the maximums. There is  a small discrepancy in the limit value of the maximum position for very long waiting times, which is underestimated by our theoretical result. The agreement can be improved by using the values of $\beta$ and $\tau$ obtained by fitting the direct relaxation function (\ref{8}) instead of their theoretical values (\ref{11}). On the other hand, it is clearly observed that the KWW prediction for the maximum position in the long waiting times regime, Eq. (\ref{19}), fails to give the correct behavior, even for moderately big values of $t_w$.

\section{Conclusions}\label{conc}

We have investigated the Kovacs effect in the one-dimensional Ising model with Glauber dynamics, in the framework of recent results derived in linear response theory. We have found an excellent agreement between the numerical and the
theoretical results, provided that the temperature jumps are such that the non-linear terms in the response of the system can be neglected. As a rule of thumb, one could say that linear response results are fine up to temperature jumps such that the relative change of the relaxation time between the initial and final temperatures is about  ten per cent in the Ising model. Then, these temperature jumps are not so small, which may make linear response relevant for actual experiments.

We have also analyzed the behavior of the position of the maximum $s_k$ and its height $K_{\max}$ as a function of the waiting time $t_w$. Simple expressions can be derived both in the limit of short and long waiting times, as compared to the characteristic relaxation time $\tau$ of the energy. The KWW function (which fits most of the relaxation of the energy at low temperatures) predicts the correct behavior of both $s_k$ and $K_{\max}$ for short waiting times but fails to do so for times comparable or larger to the average relaxation time. The maximum position $s_k/\tau$ exhibits a much slower increase with the dimensionless waiting time $t_w/\tau$ than the KWW prediction, which leads to a logarithmic divergence. Interestingly, it can be shown that $s_k/\tau$ remains bounded if the true asymptotic behavior of the relaxation function for long times, an exponential with algebraic corrections, is taken into account. The bound so obtained has been shown to fully agree with the numerical results.

These results for the position and the maximum of the Kovacs hump have been derived here for the Ising model with Glauber dynamics. However, the obtained expressions can be readily extended to any model in which (i) the overall relaxation is well fitted by an stretched exponential function (ii) the true asymptotic behavior for long times is given by an exponential, maybe with algebraic  corrections. In this sense, the behavior of the Kovacs hump should be useful to discern the true asymptotic behavior in models for which an analytical expression of the relaxation function is not known. As pointed out by Zwanzig \cite{Zw85}, it is very difficult to measure the normalized relaxation function for very long times. On the other hand, the tendency  to saturate of the position of the maximum (instead of the logarithmic divergence predicted by the KWW function) is already apparent for moderate values of the waiting time.

The results presented here improve our understanding of the Kovacs effect, by showing that it can be actually measured in the linear response regime. We have done so in one of the simplest models displaying the key behaviors of glassy systems. Furthermore, we have shown that linear response still gives a more than reasonable description of the effect for quite large temperature jumps. In particular, the position of the maximum is very well estimated by the linear theory. Therefore, the relation between the divergent (or not) character of the maximum position with the waiting time and the actual asymptotic behavior of the direct relaxation function is a robust result, which may also be possible to check in experiments. This would help to clarify the long-debated question of the true asymtpotic behavior of the relaxation function in glassy systems \cite{Zw85,PSEyA84}.

\acknowledgments
This work has been supported by the Spanish Ministerio de Econom\'\i a y Competitividad grant FIS2011-28838-C02-01 and by the
Autonomous Region of Madrid grant P2009/ENE-1597 (HYSYCOMB) (MRG) and FIS2011-24460 (AP). AP would also like to thank the Spanish Ministerio de Educaci\'on, Cultura y Deporte mobility grant PRX12/00362 that funded his stay at the Universit\'e Paris-Sud in summer 2013, during which this work has been finished.


\begin{thebibliography}{99}
\bibitem{poly} The original paper by Kovacs dealt with the relaxation of the volume of the polymer polyvinyl
acetate, but the same qualitative behavior is expected for any physical property.
\end{thebibliography}
\end{document}